# Neurogenetics of the human adenosine receptor genes: Genetic structures and involvement in brain diseases

Vincent Huin, MD, PhD[a,b], Claire-Marie Dhaenens, PharmD, PhD[a,b], Mégane Homa, MSc[a], Kévin Carvalho, MSc[a], Luc Buée, PhD[a], Bernard Sablonnière, MD, PhD[a,b]

[a] Univ. Lille, Inserm, CHU Lille, UMR-S 1172 - JPArc - Centre de Recherche Jean-Pierre AUBERT Neurosciences et Cancer, F-59000, Lille, France

[b] CHU Lille, Institut de Biochimie et Biologie moléculaire, Centre de Biologie Pathologie et Génétique, F-59000, Lille, France

**Running title:** Neurogenetics of adenosine receptors

**Correspondence to:**

Vincent Huin: vincent.huin@inserm.fr

University of Lille, Inserm UMR-S 1172, JPArc, rue Polonovski, F-59045, Lille, France.

Telephone: +33 (0)3 59 89 96 05 / Fax: +33 (0)3 20 53 85 62

**ORCID ID:**

Vincent Huin=https://orcid.org/0000-0001-8201-5406

Claire-Marie Dhaenens=https://orcid.org/0000-0003-4719-4124

Mégane Homa=https://orcid.org/0000-0002-3904-6888

Kévin Carvalho=https://orcid.org/0000-0002-6432-7622

Luc Buée=https://orcid.org/0000-0002-6261-4230

Bernard Sablonnière=https://orcid.org/0000-0003-0384-4076

Total word count of the manuscript: 6550

(excluding title, abstract, authors and affiliations, keywords, figure legends, tables, acknowledgements, authorship confirmation statement, author disclosure statements, acknowledgments, footnotes, or references)

Character count for the title: 105, 14 words

Character count for the running title, including spaces: 36

Total word count of the abstract: 168

Number of references: 136

Number of tables: 2

Number of figures: 1

Number of supplemental figures: 0

Number of supplemental tables: 0

Number of supplemental data: 1 (supplemental material)

## Authorship Confirmation Statement

### Authors' contributions:

VH wrote the manuscript.

CMD, MH, KC and BS revised the manuscript for intellectual content.

LB provided intellectual input and editorial suggestions.

### Confirmation statement:

All of the authors have reviewed and approved the manuscript prior to submission.

We confirm that the manuscript has been submitted solely to this journal and is not published, in press, or submitted elsewhere.

## Abstract

Adenosine receptors are G-protein-coupled receptors involved in a wide range of physiological and pathological phenomena in most mammalian systems. All four receptors are widely expressed in the central nervous system, where they modulate neurotransmitter release and neuronal plasticity. A large number of gene association studies have shown that common genetic variants of the adenosine receptors (encoded by the *ADORA1*, *ADORA2A*, *ADORA2B* and *ADORA3* genes) have a neuroprotective or neurodegenerative role in neurologic/psychiatric diseases. New genetic studies of rare variants and few novel associations with depression or epilepsy subtypes have recently been reported. Here, we review the literature on the genetics of adenosine receptors in neurologic and/or psychiatric diseases in humans, and discuss perspectives for further genetic research. We also provide an update on the genetic structures of the four human adenosine receptor genes and their regulation - a topic that has not been extensively addressed. Our review emphasizes the importance of (i) better characterizing the genetics of adenosine receptor genes and (ii) understanding how these genes are regulated.

## Introduction

Adenosine exerts a variety of physiological effects in most mammalian systems through four subtypes of G-protein-coupled adenosine receptors ($A_1R$, $A_{2A}R$, $A_{2B}R$, and $A_3R$), each of which is encoded by their own gene (*ADORA1*, *ADORA2A*, *ADORA2B,* and *ADORA3*, respectively). The four adenosine receptors are widely expressed in the central nervous system (CNS) in general and in neurons, glial cells and endothelial cells in particular.[1,2] In the CNS, $A_1R$ and $A_{2A}R$ modulate synaptic neurotransmission through their action on the release of various neurotransmitters (such as glutamate, γ-aminobutyric acid, acetylcholine and dopamine) and on the activation of postsynaptic receptors. Excitatory transmission is inhibited by $A_1R$ activation and stimulated by $A_{2A}R$ activation. The interplay between these two receptors helps to fine-tune synaptic plasticity.[3] In several regions of the brain, the two receptors are co-expressed and can form heteromers together[4] or with other G-coupled receptors, such as dopamine receptors.[5,6] Neuroprotective and neurodegenerative effects of $A_1R$ and $A_{2A}R$ have been observed in various neurologic diseases (Parkinson's disease, Huntington's disease, migraine, stroke, and epilepsy) and psychiatric diseases (schizophrenia, panic disorders, depression, and autism).[7,8] $A_{2B}R$ is widely expressed in many tissues, and is involved in ischemia-reperfusion injury, fibrosis, cancer, inflammation, and the immune system.[9] Lastly, $A_3R$ is involved in the modulation of ischemic diseases, inflammatory diseases, cancer, autoimmune diseases, and chronic neuropathic pain.[10] Until recently, $A_{2B}R$ and $A_3R$ were considered to have a minor physiological role because they have substantially lower affinities for adenosine and thus require higher extracellular adenosine concentrations for maximal activation. Only two studies have assessed them as genetic factors in human neurologic diseases.[11,12]

Various aspects of adenosine receptor research and the receptors' roles in the pathophysiology of the neural system, widely discussed in the current literature,[8] lead to numerous publications assessing the role of genetic variants in adenosine receptor genes and neurological or psychiatric diseases. Recent advances in these fields (including new topics with the studies of rare genetic variants[12,13] and new associations with epilepsy and depression[14,15] highlight the importance of genetic variants in adenosine receptor gene in CNS disorders.

Previous reviews or book chapters focused on the genes structure,[16] on neurobiology,[17] or on one specific gene and its neuroprotective/neurodegenerative effects.[18,19] However, these previous works do not integrate and discuss all the genetic associations reported so far. It remains difficult to have a critical overview of the neurogenetics of adenosine receptor genes with the heterogeneity of the genetic studies (in diseases, sample size, methodologies and level of evidence). This prompted us to review the current state of knowledge regarding genetic associations of variants in adenosine receptor genes with human brain diseases and to confront them to functional assessment of the variants.

The criteria and databases used for this literature search are given as supplementary material.

## ***ADORA1***

### Gene structure (Figure 1.A)

The human *ADORA1* gene was cloned in 1992 from a hippocampal cDNA library.[20,21] It spans 77 kb, and is located on chromosome 1q32.1.[22] The gene comprises six exons that undergo alternative splicing and alternative promoter usage.[23,24] The Genotype-Tissue Expression (GTEx) database[25] indicates that *ADORA1* mRNAs are found in all tissues, with the highest expression levels in the spinal cord, brain (the frontal cortex, substantia nigra, and all the basal ganglia) and testis.

According to the GENCODE database,[26] the human *ADORA1* gene encodes six different mRNAs (Figure 1.A). Three transcripts with different 5' untranslated regions (UTR) share the same coding exons (5 and 6), and yield a full-length 326-aa protein. The fourth transcript does not include exon 5, but contains an open reading frame (ORF) and might result in an N-terminal-truncated 210-aa protein. The two last transcripts contain exon 6 with an alternative 3' splice site, and are thought to encode a C-terminal-truncated 125-aa protein. Ren & Stiles showed that transcripts including exon 4 were widely expressed, whereas transcripts with exon 3 were present in the testis, kidney, fetal brain, and the frontal cortex and cerebellum of the adult brain but not in fat, skeletal muscle or heart.[23] Two alternative promoters with nonclassical TATA boxes have been found upstream of exons 3 and 4.[24,27] The promoter upstream of exon 3 is constitutively expressed at low levels, and contains a binding site for nuclear factor-kappa B (NF-κB),[28] serum response factor and activator protein 1.[29] The promoter upstream of exon 4 is more active but shows tissue-specific regulation and contains a site for an AGG element binding protein.[30] It is noteworthy that *ADORA1* has only one upstream 262 bp CpG island, which overlaps with exon 4. Interestingly, there are two upstream AUG codons in exon 4, both of which are crucial for mRNA processing and translation. Indeed, mutations in these AUG codons are associated with greater levels of translation.[24] We hypothesized that an upstream ORF (uORF) exerts post-transcriptional control on *ADORA1* transcripts including exon 4 in the same way that uORF5 regulates *ADORA2A* expression.[31]

Lastly, the ENCODE chromatin state segmentation from the University of California Santa Cruz Genomics Institute (UCSC) genome browser suggests that a third promoter is located upstream of exon 1.[32]

## Genetic risk factor in human diseases

### Psychiatric diseases

In 2009, Gotoh et al. published a case control study including 200 schizophrenic patients and 210 controls. The researchers looked for an association between 29 single-nucleotide polymorphism (SNPs) in *ADORA1* and the presence of schizophrenia in a Japanese population.[33] None of the individual SNPs was significantly associated with the disease and the small number of samples did not allow the identification of linkage disequilibrium blocks. However, a multivariate analysis evidenced an association with three haplotypes built from two or three SNPs. An important limitation concerned the control population who did not perfectly matched to the schizophrenic population. Moreover, our literature database search did not found a replication study in other Japanese population.

A case-control study focused on the association between *ADORA1* and *ADORA2A* SNPs and Tourette syndrome including 162 patients and 270 controls.[34] Patients with a GG or GT genotype at rs2228079 were underrepresented. This silent SNP (located in exon 5) was associated with earlier age at tic onset and a greater prevalence of comorbidities (the concomitant presence of depression or obsessive compulsive disorders). However these comorbidities concerned only a small number of patients with particular psychiatric disorder. Moreover, the data about the earlier age at tic onset and comorbidities in some adult patients were retrospective.

In a study of 806 Caucasian Americans, Clark et al. compared the number of alcohol dependence symptoms with the presence of variants in 71 candidate genes (including the coding, flanking and regulatory regions).[13] Although the researchers did not find a significant association with the common variants considered in isolation, alcohol dependence was linked to an aggregate of eight variants (five rare variants and three common) within intron 5 of the *ADORA1* gene. This study focused on common and rare variants and their aggregation on putative regulatory regions in the genes is original, but lacks functional validation and still need to be replicated.

These three studies lack functional data and are balanced by negative results provided similar case-control studies evidencing no genetic association between *ADORA1* SNPs and bipolar affective disorders[35] or methamphetamine dependence or psychosis.[36] Moreover, the role of *ADORA1* common variant in anxiety disorders does not seem to be important regarding the study of Rogers et al. who did not found an association between seven *ADORA1* SNPs and effects of caffeine (an antagonist of $A_1R$) on anxiety.[37] Lastly, recent genome-wide association studies (GWAS) on schizophrenia,[38,39] Tourette syndrome[40] and alcohol dependence[41] did not replicate these associations. The genetic associations of *ADORA1* common variants in psychiatric diseases remain controversial and should then be interpreted with caution.

**Neurologic and neurodegenerative disorders**

*ADORA1* mutations are thought to have a causal role in familial Parkinson's disease.[42] In a study of a large consanguineous family in which two members were affected by early-onset Parkinsonism and cognitive dysfunction, Jaberi et al. used homozygosity mapping and exome sequencing to detect a homozygous missense mutation c.835G>A, p.(Gly279Ser) in exon 6 of the *ADORA1* gene. Segregation analyses, bioinformatics predictions and the variant's rarity made this missense a good candidate. However, the Gly279Ser variant did not have a functional effect on the interaction between $A_1R$ and the dopamine D1 receptor. Based on these genetic studies, and given $A_1R$'s role in neurodegenerative diseases, Jaberi et al. hypothesized that *ADORA1* was a reasonable candidate disease-causing gene. Moreover, the researchers suggested that *ADORA1* was located within the Parkinson's disease locus PARK16, which has been identified in several populations in earlier GWAS.[43,44] However, it was later demonstrated that *ADORA1* was not in the PARK16 locus, and that mutations in *ADORA1* could not drive the PARK16 association signal.[45] Moreover, Jaberi et al. identified another homozygous variant in *PTRHD1* that was later reported as a more likely candidate disease-causing gene. Indeed, Khodadadi et al. reported a different mutation in *PTRHD1* in their study of another Iranian family with a similar clinical presentation.[46,47] *ADORA1* variants are not a common cause of Parkinson's disease or dementia with Lewy bodies[45] and the role of the Gly279Ser variant in early-onset Parkinsonism with cognitive dysfunction is subject to debate.

Variants in *ADORA1* are associated with susceptibility to epilepsy after severe traumatic brain injury.[48] In a retrospective study of 206 patients with severe traumatic brain injury (excluding penetrating head injury), Wagner et al. genotyped 6 SNPs in *ADORA1*. There was an association between the AA genotype at rs3766553 and an elevated incidence of early post-traumatic seizures, whereas the GG genotype at rs3766553 and the CT genotype at rs10920573 were associated with an elevated incidence of late-onset seizures. However, the study was not informative regarding the risk to develop recurrent seizures.

To conclude, whereas the role of *ADORA1* in rare Mendelian cause of Parkinsonism has been questioned if not refuted; the role of *ADORA1* common variants in epilepsy after traumatic brain injury seems promising and might lead up to a better risk stratification and maybe to the development of new therapeutic strategies.

## ***ADORA2A***

### Gene structure (Figure 1.B)

Of the four human adenosine receptor genes, *ADORA2A* has the most complex structure. It was first cloned from a hippocampal library in 1992.[49] It is 24 kb long, and is located on chromosome 22q11.23.[50–52] Gene expression for *ADORA2A* mRNA has been evidenced in many tissues, including the liver, heart, lung, immune system, and brain.[53] In the GTEx project,[25] *ADORA2A* mRNAs were found to be expressed in all tissues; the highest expression levels were observed in the basal ganglia (the nucleus accumbens, putamen, and caudate), blood, spleen, lung, and cerebellum.

*ADORA2A* is composed of two invariant coding exons and several alterative exons involved in the formation of a 5'-UTR. Using molecular and bioinformatics analyses, Yu et al. identified new exons encoding the 5'-UTR of *ADORA2A* in the mouse and in the human.[54] In the human, the researchers evidenced six variants of non-coding exon 1, and referred to the invariant coding exons as exons 2 and 3. All transcripts encoded the same protein but displayed tissue-specific expression patterns. Kerth et al. identified eight 5'-UTR splice variants of *ADORA2A* produced by the alternative transcription and/or splicing of five noncoding exons upstream of the two invariant coding exons.[55] In the Ensembl database, *ADORA2A* mRNAs have eight different transcription start sites (TSSs) (Figure 1.B), and non-coding RNAs give rise to three other TSSs. Not all the transcripts have been fully sequenced, and levels of evidence are low for some. In the GENCODE database,[26] *ADORA2A* has four transcripts with an alternative 5'-UTR encoding the same 412-aa protein (Figure 1.B).

Accordingly, the several putative promoter regions have not yet been extensively characterized. Most studies to date have been performed in the rat[56–59] or the mouse.[54] A few researchers have sought to characterize the human promoters[60,61], mostly by focusing on a 2000 bp region of upstream of exon 2. By considering the histone code and the chromatin state segmentation data[62] from the ENCODE project,[63] we observed five non-overlapping promoter regions. As determined by the CpG Islands Track Settings tool (UCSC), only one promoter region comprises a 1272 bp CpG island encompassing different alternative first non-coding exons (Figure 1.B). Methylation of this CpG island and the surrounding CpG shore was assessed in human brain samples from healthy individuals,[64–66] patients with Huntington's disease (HD),[67] and schizophrenic patients[68] - demonstrating that *ADORA2A*'s promoters are under epigenetic control. It is very likely that each promoter region has different transcription factor binding sites; these might control the expression of the *ADORA2A* gene under various conditions (e.g. as a function of the cell type, the development stage and/or the response to different stimuli).

This complexity is heightened by the existence of different 5'-UTRs, which might recruit different RNA-binding proteins and thus regulates the degradation, storage, and/or translation of the corresponding *ADORA2A* mRNAs.[69] In the rat, the different 5'-UTRs can repress translation via the use of a conserved ORF (notably in humans) located 65 nucleotides upstream of the usual translation initiation site.[70] Moreover, the rat uORF leads to the translation of a 134-aa protein called uORF5.[71] Hence, the *ADORA2A* gene is a rare example of a human dual-coding gene.

Yu et al. reported on a novel transcript that reads in the opposite direction to the *ADORA2A* gene.[54] Indeed, there appears to be an antisense non-coding RNA gene (*ADORA2A-AS1*) at the same locus. Three isoforms of *ADORA2A-AS1* have at least one exon that complements part of the *ADORA2A* UTR. Lastly, read-through transcription between the upstream *SPECC1L* gene (coding for sperm antigen with calponin homology and coiled-coil domains 1-like) and *ADORA2A* has been reported. This *SPECC1L-ADORA2A* transcript results in a non-coding RNA that probably induces nonsense-mediated mRNA decay and is then unlikely to produce a protein.[72] The functions of *SPECC1L-ADORA2A* and *ADORA2A-AS1* are unknown but as other long non-coding RNAs (lncRNA), they might regulate different genes in the same locus, including *ADORA2A*. Indeed, the lncRNAs are known as regulator of gene expression modulating the transcription, splicing, and translation of other mRNA by RNA-DNA or RNA-RNA interaction.[73] They have been implicated in numerous psychiatric/neurologic diseases, in neurodevelopment and neurodegeneration.[74,75] Then, genetic variants in these non-coding genes can (i) be in linkage disequilibrium with other more relevant variant in *ADORA2A*, (ii) be also located in an important region for the regulation of *ADORA2A*, like an enhancer and then modulate its transcription, but it is likely that (iii) some genetic variants affecting the expression of *SPECC1L-ADORA2A* or *ADORA2A-AS1* could indirectly modify the expression of genes regulated by these lncRNAs. For example, *ADORA2A-AS1* might form RNA duplexes with *ADORA2A* mRNA whereas *SPECC1L-ADORA2A* could induce epigenetic changes in *ADORA2A*.

## Genetic risk factor in human diseases

### Psychiatric diseases

From 1998 onwards, *ADORA2A* variants were suspected to confer susceptibility to various anxiety disorders (e.g. panic disorder).[76] In an initial comparison between 89 German patients with panic disorder and matched controls, researchers observed a significant association of the disease with the T allele and the TT genotype of a silent SNP (rs5751876) located in *ADORA2A* exon 3. Later, Hamilton et al. highlighted a genetic linkage between panic disorder (especially the form with agoraphobia) and a haplotype encompassing rs5751876.[77] All the significantly associated haplotypes comprised the C allele of rs5751876; this contradicted Deckert et al.'s report that the rs5751876 T allele was associated with an elevated risk of anxiety disorder. This discrepancy suggests that rs5751876 is not a causal variant conferring an elevated risk of anxiety disorder. Furthermore, it has also been reported that rs5751876 is associated with (i) various anxiety-related personality types[76] and (ii) psychophysiological markers of anxiety-related arousal of the sympathetic nervous system in blood-injury phobia.[78] The same research group replicated the association between rs5751876 (the TT genotype) and panic disorder.[79] Again, the researchers found a stronger association in the agoraphobia subgroup - prompting them to suggest that the association with the panic disorder population as a whole might originate in this subgroup alone. Hohoff et al. found an association between an anxious personality and a haplotype composed of eight SNPs in *ADORA2A* (including rs2298383 and rs5751876). In one study rs2298383 was not associated with panic disorders in the absence of agoraphobia.[80] However, a *post hoc* power analysis showed that the study including 71 patients and 100 controls was not powered to reliably detect a true association. Studies of different anxiety arousal paradigms in healthy subjects (such as caffeine-induced anxiety[81,82] and anxiety response to amphetamine[83]) also highlighted an association with *ADORA2A* variants. Recent studies have highlighted a complex interaction between *ADORA2A* genetic variants and anxiety/attentional responses linked to caffeine.[37,84–86] Taken as a whole, these data provide strong evidence for a role of the *ADORA2* gene in anxiety disorders, and fit with the abnormally high level of anxiety in *Adora2a*-knockout mice.[87] Most of these studies were conducted in Caucasian cohorts, and the association between rs5751876 and anxiety disorders was not replicated in two case control studies conducted in Chinese[88] or in Japanese[89] populations (respectively, 104 cases *versus* 192 controls; and 91 cases *versus* 100 controls) - suggesting that the association is ethnicity-dependent. These findings also suggest that rs5751876 (the most widely studied *ADORA2A* polymorphism) is not the causal variant conferring an elevated risk of anxiety; in fact, it is probably a tag SNP for the true causal variant. The latter might therefore be less frequent and/or not in linkage disequilibrium with rs5751876 in Asian populations. Another hypothesis is that *ADORA2A* confers susceptibility to anxiety in some patient subgroups (e.g. patients with agoraphobia).[77,79] Another example was provided in 2012 by Subaran et al., who described an association with rs5751876 and rs4822492 in a subset of patients with panic disorders and bladder syndrome ($n$=92) but not in the whole cohort of patients with panic disorder ($n$=351).[90]

Molero et al. studied the relationship between six *ADORA2A* polymorphisms in strong linkage disequilibrium and attention-deficit hyperactivity disorder traits.[91] After correction for multiple comparisons, the rs35320474 SNP was still significantly associated with inattention symptoms but not with hyperactivity-impulsivity. Very recently, Fraporti et al. confirmed the association between the TT genotype of rs2298383 and the presence of anxiety disorders in children with attention-deficit hyperactivity disorder.[92]

Janik et al. reported a link between *ADORA1*, *ADORA2A* and Tourette syndrome.[34] The T allele of rs5751876 was overrepresented among patients, and was associated with earlier age at onset. However, in contrast to previous studies, the T allele was underrepresented in patients with concomitant attention-deficit hyperactivity disorder and patients with anxiety but not obsessive-compulsive disorder. The later result was more questionable as it was obtained on small samples including only 128 non anxious patients with Tourette syndrome versus 34 anxious patients.

Freitag et al. linked several *ADORA2A* SNPs to autism spectrum disorders (ASD) in a genetic case-control analysis of 98 patients and 234 controls.[93] The CC genotype of rs2236624 (located in intron 3) was overrepresented in this group of patients. Increased current social interaction, nonverbal communication and repetitive behavior scores were associated with rs3761422 (CT and TT genotypes), rs5751876 (CT and TT genotypes), and rs35320474 (T/del and del/del genotypes). Lastly, the researchers evidenced a significant association between a haplotype of eight SNPs (including rs5751876) and anxiety in the group of 98 patients with ASD. The most frequent haplotype ACTCCCTG (built with the rs5751862, rs5760405, rs2298383, rs3761422, rs2236624, rs5751876, rs35320474 and rs4822492 variants) was carried by 73.3% of the patients with severe anxiety disorders but only 33.9% of the patients without a history anxiety. However, it should be noted that although the allele C of rs5751876 was significantly less frequent in autism patients, it was overrepresented in the subset of individuals with a history of anxiety.

The genetic link between *ADORA2A* and psychosis is more subject to debate. In a study of small number of schizophrenic patients (*n*=18) and healthy controls (*n*=33), Jagannathan et al. evaluated the relationship between gray matter volume (as measured using structural MRI) and 367 different SNPs in candidate genes.[94] The researchers found an inverse correlation between a lack of gray matter in the frontal, temporal and thalamic regions (previously reported to be altered in schizophrenia) and 18 SNPs in 16 genes, including rs3761422 in *ADORA2A*. However, it had previously been reported that *ADORA2A* variants (including rs5751876) were not involved in the development of schizophrenia.[95–97] The absence of a relationship was confirmed by two recent GWASs of patients with schizophrenia.[38,39]

Recently, Oliveira et al. published the results of a cross-sectional population-based study of 1253 participants, of whom 228 were experiencing current depression.[15] The TT genotype of rs2298383 was less frequent in patients experiencing current depression than in controls. After adjusting for confounding variables, this association remained significant (odds ratio (OR) [95% confidence interval (CI)]=0.631 [0.425–0.937]; $p$=0.022). The TT genotype was associated with protection against sleep dysfunction and attentional impairment - two common symptoms observed in patients with depression. However, an earlier case-control study performed in an Asian population and including 192 mood disorder patients and 216 controls did not evidence an association between rs5751876 and major depression or bipolar disorder.[98]

**Neurologic and neurodegenerative disorders**

In 2007, *ADORA2A* was linked to migraine with aura.[99] A haplotype block spanning the rs5760405, rs5751876, rs35320474, rs5751862, rs3761422 and rs4822492 SNPs in *ADORA2A* was tested in 265 patients with migraine (of whom 122 had migraine with aura) and 154 controls. A GCCCTG haplotype was more frequent in patients suffering from migraine with aura than in controls. However, there was no significant difference between patients with migraine without aura and controls.

Genetic variants in the *ADORA2A* gene are known to be genetic modifiers in the pathogenesis of HD, a hereditary neurodegenerative disorder caused by the expansion of CAG repeats within the *HTT* gene. The broad range of ages at onset is mostly explained by the number of CAG repeats but is also related to various environmental and genetic factors - including *ADORA2A* polymorphisms. A study published in 2009 reported an earlier age at onset (by 3.8 years) in patients with the TT genotype at rs5751876 than in patients with the CC genotype ($p$=0.02).[100] The impact of *ADORA2A* genotype on residual variation was modest; however, these results have been replicated in two other studies.[101,102] One of the latter found that rs2298383 in intron 1 (in linkage disequilibrium with rs5751876) to be the most strongly associated *ADORA2A* variant.[101] These findings are supported by our knowledge of other environmental modifiers of HD (such as caffeine intake[103]) and data from animal models on the roles of $A_1R$ and $A_{2A}R$.[104]

Horgusluoglu-Moloch et al. used the dataset from the Alzheimer's Disease Neuroimaging Initiative cohort and selected 1,563 non-Hispanic Caucasian participants to correlate different quantitative phenotypes (including hippocampal volume, metabolic activity, cerebrospinal fluid total tau level, amyloidosis in the hippocampus, and composite memory) with 18407 SNPs in 81 genes from a targeted neurogenesis pathway.[105] The researchers showed that participants with at least one copy of the minor allele T of rs9608282 (located in the first intron of *SPECC1L*, upstream of *ADORA2A*) had a greater mean hippocampal volume and a greater neurogenesis-related hippocampal sub-region volume than participants bearing the other genotype. Horgusluoglu-Moloch et al. replicated these results in a meta-analysis that included two other cohorts (adding 277 additional participants) and evidenced an association with better episodic memory performance, and with a low cerebrospinal fluid total tau level. These results suggest that *ADORA2A* has a role in neurogenesis in adults. However, an analysis of the BRAINEAC brain tissue microarray-based gene expression database (http://www.braineac.org/) found that rs9608282 is not associated with *ADORA2A* expression levels in the hippocampus. Future research will have to validate this *SPECC1L* SNP and assess its potential function in the adenosine-related pathway.

Shinohara et al. identified an *ADORA2A* haplotype “A” (built from a C allele at rs2298383, a T allele at rs5751876, a deletion at rs35320474 and a C allele at rs4822492) as a risk factor for a severe form of encephalopathy in children following severe febrile seizures: acute encephalopathy with biphasic seizures and late reduced diffusion.[14] The haplotype A was significantly more frequent in patients ($n$=85) than in controls ($n$=184). The most significant association was obtained in a recessive model (OR [95%CI]=2.32 [1.32-4.08]; $p$=0.003). The level of *ADORA2A* mRNA was 1.4-fold higher in the brain tissue of individuals with a homozygous haplotype A (*i.e.* diplotype AA), relative to individuals with AB or BB diplotypes. This finding was confirmed by the mRNA and protein expression levels measured in lymphoblasts. Lastly, the researchers evidenced greater cAMP accumulation in response to adenosine in lymphoblasts with the AA diplotype than in cells with the BB diplotype ($p$=0.0006). These results support the hypothesis whereby elevated expression of *ADORA2A* in children may alter the intracellular adenosine/cAMP signal cascade, enhance excitatory neurotransmitter release, and thus favor excitotoxicity.

**Neurologic traits**

Two studies have suggested that a TT genotype at rs5751876 is associated with greater sensitivity to sensory stimuli. Firstly, *ADORA2A* variants were associated with differences in the efficiency of pre-attentive sensory memory subprocesses.[106] The researchers examined the results of a memory task after a visual stimulus in 199 healthy participants. Although the initial visual availability of stimulus information was greater in participants carrying the rarer homozygous genotype (*i.e.* TT genotypes for rs5751876 and rs2298383), less information was transferred to the working memory. This finding suggests that pre-attentive sensory memory subprocesses are modulated by *ADORA2A*. Secondly, Geiger et al. correlated functional MRI and genotyping results in healthy human subjects.[107] A TT genotype at rs5751876 (associated with an elevated anxiety risk) was found to modulate a fronto-insular network involved in attention and in the processing of interoceptive and exteroceptive information; it was suggested that this modulation increases the likelihood of anxiety and anxiety disorders.

In a pilot study of 19 healthy volunteers, Rupp et al. assessed the relationship between sleep restriction resistance and *ADORA2*A and *PER3* genotypes.[108] An

rs5751876 TT genotype group (*n*=9) showed lower behavioral resiliency after sleep restriction than a group with the CT genotype (*n*=9). Several studies has confirmed the relationship between *ADORA2A* variants and sleep, with notably an association between genetic variants in *ADORA2A* (especially rs5751876) and individual sensitivity to sleep disturbance caused by caffeine consumption.[109–112]

To conclude, the genetic variants in *ADORA2A* are the most studied among adenosine gene receptor. They are associated with a lot of CNS diseases or traits, but sometimes provided contradictory results.[76,77] The most likely genetic associations concern HD, epilepsy, depression and anxiety disorders because of functional assessment of the variants, prospective design and/or studies on larger samples. However, efforts are needed to identify the true functional variant(s).

## ***ADORA2B***

### Gene structure (Figure 1.C)

The human *ADORA2B* gene was cloned from a human hippocampal cDNA library in 1992.[113] It spans 31 kb, and is located on chromosome 17p12.[114,115] Although only a single copy of *ADORA2B* is present in the genome, a pseudogene (adenosine A2B receptor pseudogene 1, *ADORA2BP1*) was found on chromosome 1q32.2.[115] *ADORA2BP1* is a processed pseudogene that lacks an intronic region. It may have been created during evolution by integration of *ADORA2B* mRNA into a chromosome break site, followed by DNA synthesis and repair. *ADORA2B* is composed of two coding exons and an intron. The gene comprises a 1756 bp CpG island encompassing the whole of exon 1 and the core promoter. The *ADORA2B* promoter contains binding sites for hypoxia-inducible factor 1-alpha,[116,117] hypoxia-inducible factor 2-alpha,[118] and Myb-related protein B.[119] According to the GTEx database,[25] *ADORA2B* mRNAs are found in all tissues, with the highest expression levels in the skin, esophagus, vagina, colon, and anterior cingulate cortex.

**Genetic risk factor in human diseases**

*ADORA2B* has been highlighted as a potential modifier gene in prodromal HD with a variation in gray matter volume.[11] After correcting for the effect of the disease burden effect (*i.e.* the number of *HTT* gene CAG repeats, and age), Liu et al. searched for genetic factors that influenced prodromal HD in a multicenter study including 715 prodromal individuals. A GWAS did not reveal an association that remained significant after false discovery rate correction. However, a separate analysis of 3,404 SNPs in 310 HD pathway genes evidenced a correlation between two SNPs in strong linkage disequilibrium (rs71358386 in *NCOR1*, and rs78804732 in *ADORA2B* intron 1) and the gray matter volume in the cuneus lobe; the presence of the minor alleles was correlated with a low gray matter volume. The major limitation consisted in the absence of correlation between SNPs and motor or cognitive function in prodromal individuals. Consequently, functional studies and longitudinal study in HD patients will be required to assess a modifier role of *ADORA2B.*

## *ADORA3*

### Gene structure (Figure 1.D)

The human ADORA3 gene was first cloned in 1992.[120,121] It is a short (4.3 kb) gene and is located on chromosome 1p13.2.[122,123] It is composed of two exons and one intron,[124] and encodes a single 318-aa protein. According to the GTEx database,[25] *ADORA3* mRNAs expression levels are highest in the testis, spinal cord, adrenal gland, subtantia nigra, and spleen.

The entire *ADORA3* gene is encompassed by the first intron of the isoforms 2 to 4 of transmembrane and immunoglobulin domain containing 3 gene (*TMIGD3*). Moreover, isoform 1 of *TMIGD3* shares its 5' terminal exon with almost all of the first exon of *ADORA3*. This alternative TSS results in a fusion product that includes the 116 N-terminal amino acids of $A_3R$. Data on *TMIGD3* are scarce, although one study revealed a role for *TMIGD3* isoform 1 as a suppressor of NF-κB and osteosarcoma progression.[125] The *ADORA3* gene's promoter region has not been extensively studied but reportedly does not contain a TATA box; however, the existence of a CCAT box is subject to debate.[126,127] Our analysis of the *ADORA3* promoter region evidenced a CAAT box located 98 bp upstream of the TSS. Functional binding of Octamer-Binding Protein 1 and NF-κB transcription factors was found at the proximal part of the *ADORA3* promoter region, and binding was inhibited by the rs1544224 polymorphism (c.-564T>C, p.?) located in the 5'-UTR (encoded by exon 1).[128]

### Genetic risk factor in human diseases

Prompted by $A_3R$'s known role in the neuromodulation of serotonin (a neurotransmitter involved in many neuropsychiatric phenotypes, including autism),

Campbell et al. examined a putative genetic link between rare coding variants of *ADORA3* and autism. Using allelic association testing in 958 families (1,649 probands; 4,150 samples), the researchers found no evidence supporting an association between a common variant at *ADORA3* locus and ASD. However, in a case-control study including 185 unrelated ASD cases and 305 controls followed by a subsequent replication analysis of exome sequence data from 339 cases and pair-matched controls,[12] two rare coding variants (Leu90Val and Val171Ile) were significantly enriched in cases *vs.* controls. They were then assessed *in vitro* for their functional impact but showed incomplete penetrance and had opposite effects on antidepressant-sensitive 5-HT transporter activity. The relationship with disease was significant in the combined analysis using whole-exome sequencing data but not in the replication analysis. Taking into account replication data, only Leu90Val remain statistically significant ($p$=0.040). In total, Leu90Val was found in 4 patients among 524 ASD cases and was absent in 644 control. The subjects in this study were not matched based on genome-wide genotype data. This limitation could constitute a bias, especially for the frequency of rare variants which can be very different between subgroups of a population with similar ethnicity. Such limitation is important regarding the small number of carriers. Although the researchers provided functional data of the two rare variants in cell culture, the association of rare *ADORA3* variants and ASD should be interpreted with caution. Further studies in larger samples are needed to confirm this genetic association.

**Which genetic variants in adenosine receptors genes are functionally relevant?**

Despite their associations in several diseases, the functional effect of variants in adenosine receptors genes has been poorly described. The three more frequently studied *ADORA2A* SNPs (*i.e.* rs5751876, rs2298383 and rs35320474) and their associations with human brain diseases or traits are summarized in Table 2. The most-studied variant (rs5751876, c.1083T>C, p.(=)) is a synonymous variant that lacks a validated functional impact on protein function or synthesis. It has been suggested that several *ADORA2A* polymorphisms in linkage disequilibrium with rs5751876 are causal variants. For example, rs35320474 (c.*458dup, p.?, located in the 3'-UTR region of exon 4) has been discussed as a variant that might change $A_{2A}R$ expression.[81] Indeed, there is no data in favor of a transcriptional effect of rs35320474 in GTEx, but it modifies a U-rich motif. As these motifs can recruit RNA-binding proteins, rs35320474 might modify the translation of *ADORA2A* mRNA. However, such functional effect has not been published for rs35320474. The second most intensively studied polymorphism in *ADORA2A* is the rs2298383 SNP (c.-3206C>T, p.?), located in a regulatory triplex-forming oligonucleotide target sequence that has been conserved in mammals and is involved in the regulation of gene expression.[129] It is also noteworthy that rs2298383 variant is located in the terminal 3' exon of one of the isoforms of *ADORA2A-AS1* and is associated with *ADORA2A-AS1* mRNA expression.[25] It is well known that antisense RNAs regulate gene expression on the opposite strand. Hence, it would be interesting to study the effect of rs2298383 on the mRNA expression of *ADORA2A* and *ADORA2A-AS1*. In line with a putative transcriptional effect, rs2298383 is also an expression quantitative locus associated with *ADORA2A* mRNA expression in diverse tissues including different regions of the brain (*i.e.* nucleus accumbens, amygdala, cortex and hippocampus) (https://gtexportal.org/home/snp/rs2298383).[25] More precisely, the TT genotype is associated with increased *ADORA2A* mRNA. But these results are in contradiction with the functional studies reported by Shinohara et al. who found elevated $A_{2A}R$ mRNA and protein expression levels and protein activity with associated *ADORA2A* haplotype A, including the C allele of rs2298383.[14] It is then probable that rs2298383, like the other variants in *ADORA2A*, only tag the true causal variant(s). In 2014, Hohoff et al. used positron emission tomography to investigate the influence of *ADORA1* and *ADORA2A* SNPs on the *in vivo* distribution and availability of cerebral $A_1Rs$ in healthy humans.[130]

Indeed, $A_{1A}Rs$ and $A_{2A}Rs$ are frequently co-expressed, and interact directly by heteromerization. Hohoff et al. observed that *ADORA2A* SNPs are associated with increased binding and availability of $A_1Rs$ in various regions of the brain. The CT and TT genotypes at rs2236624 are associated with greater distribution and availability of $A_1R$ in the superior frontal gyrus and dorsolateral prefrontal cortex, whereas the CT and TT genotypes at rs5751876 are associated with greater distribution and availability in the entorhinal cortex, hippocampus, and cortex as a whole. The brain regions associated with *ADORA2A* SNPs are involved in emotion processing, which might contribute to their effects on anxiety and panic disorder. The SNPs rs10920568 and rs3766566 in *ADORA1* also correlated with the *in vivo* distribution and availability of cerebral $A_1Rs$.

None of *ADORA1* common variants rs3766553 and rs10920573 (associated with epilepsy after traumatic brain injury[48]) has been assessed during functional studies. We noted that rs10920573, located in intron 5, reported in the GTEx database,[25] is associated with variation in expression of *ADORA1* mRNA in the whole blood ($p$=8.8 $e^{-15}$). Rarer allele T, and genotype TT of rs10920573 are associated with an increase of *ADORA1* mRNA, but Wagner et al. found that it was the heterozygotes genotype CT that was at greatest risk of epilepsy which is not in favor of protective effect of either allele. Functional impact of the two SNPs reported by Liu et al. is not established and we even don't know which gene between *ADORA2B* and *NCOR1* is implied in prodromal HD.[11] *NCOR1* and *ADORA2A* (but not *ADORA2B*) are part of the HD pathway, but it is noteworthy that SNP rs71358386 in *NCOR1* regulates expression of *ADORA2B* (and not *NCOR1)* in various tissues including the brain cortex.[25] Minor allele G of rs71358386 is negatively correlated to grey matter volume in the cuneus lobe and is associated with low *ADORA2B* mRNA.[25] How a decrease in *ADORA2B* mRNA (and then probably in $A_{2B}R$ too) induces a decrease in grey matter and potentially favors HD progression is not fully understood. However, recent studies demonstrated the existence of brain $A_{2A}R$-$A_{2B}R$ heteroreceptor complexes, that might inhibit neurodegeneration in models of brain diseases by $A_{2A}R$ inactivation.[131] The functional effect of the rare coding variant Leu90Val in *ADORA3* was demonstrated in cell culture by Campbell et al. The researchers showed elevated basal cGMP levels and response to an $A_{3A}R$ agonist compared with the wild type receptor. Moreover, the cells transfected with *ADORA3* and 5-HT transporter gene showed an increase in antidepressant-sensitive 5-HT transporter activity.[12]

**Conclusions**

The results of a large number of genetic studies suggest that variants in adenosine receptor genes are associated with neurologic or psychiatric diseases. However, most of the research performed to date has been based on small case-control studies. Without any functional evidence and replication in other populations, such studies do not provide enough evidences and must be interpreted with cautions because smaller sample size increase the risk that observations will be due to chance. The multiplication of genetic studies of recurrent genetic variants in multiple disorders, as seen for rs5751876 in psychiatric disorders, increases the risk of false-positive too. It may explain the heterogeneity of the findings and for example why the incriminated allele of rs5751876 associated with panic disorder varies depending the studies[76,77] or why *ADORA2A* is absent in larger studies like GWAS on anxiety disorder.[132] However, there are too many genetic associations between *ADORA2A* SNPs and anxiety disorders to disregard them all. Moreover, a growing body of evidence suggested that $A_{2A}R$ influence the fear network[133] and bring rational explanations to these genetic associations. This is in line with the recent report of increased $A_{1A}R$ availability in regions of the fear network, in *ADORA2A* risk allele carriers.[130]

The development of new study designs in genetic research has prompted renewed interest in specific phenotypes,[14] cross-sectional population-based analysis[15] and rare variants.[12,13] The study of rare variant is of particular interest as it is based on the common disease-rare variant hypothesis, according an increasing role of rare variant in human complex diseases or traits.[134] A screen for rare variants and/or the aggregation of variants in regulatory regions could also be interesting for *ADORA2A* because none of the SNPs reported to date have proved to be functional and may only be tags for one several rare variants in linkage disequilibrium. Such rare variants restricted to smaller groups of populations could also explain the ethnicity-dependent and/or heterogeneity of the findings concerning *ADORA2A* SNPs.

In the future, genetic research should take account of a number of confounding factors (e.g. ethnicity and gender) and common gene-environment interactions (e.g.

caffeine intake). Further studies are particularly needed to integrate the role of caffeine, a widely consumed antagonist of adenosine receptors, in HD,[103] depression[135] and neuropsychiatric disorders.[18]

Another major task relates to the characterization of the genes' structure and regulatory mechanisms. Indeed, most of the structural specificities and genetic variants in the adenosine receptor genes do not lead to changes in the protein sequences or biochemical properties. However, adenosine receptor expression is frequently deregulated in a context of neurodegenerative disease. Thus, there is great interest in the role of adenosine receptor genes and their posttranslational regulation as it could be the key to understand the deregulations of adenosine receptors in many common neurodegenerative diseases.

Lastly, there is a growing body of evidence suggesting a modulation of the neurodegenerative effects of $A_{2A}R$ through its interaction with other adenosine receptors.[131] Further studies should include all robust associations in adenosine receptors genes, in related non-coding RNAs and in other relevant genes. For example, it would be interesting to study together variants in *ADORA2A* and in different genes implicated in adenosine metabolism or nucleoside transporter, such as the variants in *SLC29A3* which are associated with altered adenosine transport and predispose to depression with disturbed sleep.[136]

## Acknowledgments

The research leading to these results received funding from the French government's LabEx program ("Development of Innovative Strategies for a Transdisciplinary approach to Alzheimer's disease"- DISTALZ), the University of Lille, and the Institut National de la Santé et de la Recherche Médicale (INSERM).

## Author disclosure statements

No competing financial interests exist.

**Table 1.** Adenosine receptor genes and proteins

| **Official gene Name*** | **Old Name** | **Reference mRNA transcript#** | **Protein name** | **Protein length (aa)** |
|---|---|---|---|---|
| *ADORA1* | *RDC7* | NM_000674.2 | $A_1R$ | 326 |
| *ADORA2A* | *RDC8* | NM_001278497.1 | $A_{2A}R$ | 412 |
| *ADORA2B* | - | NM_000676.2 | $A_{2B}R$ | 332 |
| *ADORA3* | - | NM_000677.3 | $A_3R$ | 318 |

* HUGO nomenclature; # the reference mRNA transcripts are the longest transcripts listed in the NCBI RefSeq database (GRCh37/Hg19) and encode the longest protein isoform.

**Table 2.** Significant associations in brain diseases or traits of rs5751876, rs2298383 and rs35320474 in *ADORA2A*

| Associated brain diseases or traits | Studied population | | | Risk allele | Risk genotype | Ref |
|---|---|---|---|---|---|---|
| | Patients | Ethnicity | Country | | | |
| **rs5751876** | | | | | | |
| Panic disorder | 99 patients and 99 controls | Caucasian | Germany | T | TT | 76 |
| Panic disorder | 70 multiplex families and 83 triad families | Caucasian* | USA | C | | 77 |
| Sympathetic measures during venipuncture | 17 patients with blood-injury phobia | Caucasian | Germany | | TT | 78 |
| Panic disorder | 457 patients and 457 controls | Caucasian | Germany | T | | 19 |
| Panic disorder and agoraphobia | 341 patients and 457 controls | Caucasian | Germany | C | | 19 |
| Anxious personality trait | 424 healthy participants | Caucasian | UK | C | | 19 |
| Self-reported anxiety after caffeine administration | 94 healthy participants | Caucasian* | USA | | TT | 81 |
| Self-reported anxiety after d-amphetamine administration | 99 healthy participants | n.a. | USA | | TT | 83 |
| Self-reported anxiety after caffeine administration | 416 healthy participants | Caucasian | UK | | TT | 37 |
| Startle reflex after caffeine administration | 110 healthy participants | Caucasian | Germany | | TT | 84 |
| Prepulse inhibition/facilitation after caffeine administration | 114 healthy participants | Caucasian | Germany | | TT | 85 |
| Decreased attention after caffeine administration | 106 healthy participants | Caucasian | Italy | | CC | 86 |
| Decreased motor executive control after caffeine administration | 106 healthy participants | Caucasian | Italy | | TT | 86 |
| Panic disorder patients with BS | 92 patients BS+ and 259 BS- | Caucasian | USA | T | | 90 |
| Tourette syndrome | 162 patients and 270 controls | Caucasian | Poland | T | | 34 |
| Tourette syndrome with concomitant ADHD | 63 patients ADHD+ and 99 patients ADHD+ | Caucasian | Poland | C | | 34 |
| Tourette syndrome with concomitant anxiety | 34 patients anxiety+ and128 patients anxiety- | Caucasian | Poland | C | | 34 |
| Increased autism-specific impairment | 98 patients | Caucasian | Germany | T | | 93 |
| Migraine with aura *vs* without aura | 122 patients aura+ and 143 aura- | Caucasian | Germany | C | | 99 |
| Earlier age of onset HD | 791 HD patients | Caucasian | France | | TT | 100 |
| Earlier age of onset HD | 419 HD patients | Caucasian | Germany | | TT | 101 |
| Earlier age of onset HD | 18 HD patients | n.a. | Uruguay | T | | 102 |
| Variation in pre-attentive sensory memory sub-processes | 199 healthy participants | Caucasian | Germany | | TT | 106 |
| AESD | 85 patients and 184 controls | Asian | Japan | T | | 14 |
| Attention and processing of information | 65 healthy participants | Caucasian | Germany | | TT | 107 |
| Lower behavioral resiliency after sleep restriction | 20 healthy participants | n.a. | USA | | TT | 108 |
| Insomnia-like EEG pattern after caffeine administration | 19 healthy participants | Caucasian | Switzerland | | CC | 109 |
| Sustained vigilant attention after prolonged wakefulness | 45 healthy participants | Caucasian | Switzerland | T | | 110 |
| Lower effect of caffeine on attention after prolonged wakefulness | 45 healthy participants | Caucasian | Switzerland | T | | 110 |
| Sensitive to caffeine's effects on sleep | 2402 healthy participants (1470 families) | n.a. | Australia | C | | 111 |
| Sleep disturbance caused by caffeine consumption | 926 healthy participants | n.a. | Brazil | T | | 112 |
| $A_{1A}R$ availability in brain regions of fear network | 28 healthy participants | Caucasian | Germany | T | | 130 |
| **rs2298383** | | | | | | |
| Anxious personality trait | 424 healthy participants | Caucasian | UK | T | | 79 |
| Self-reported anxiety after caffeine administration | 102 healthy participants | Caucasian* | USA | | CC | 82 |
| Anxiety disorder in ADHD patients | 478 ADHD patients | Caucasian* | Brazil | | TT | 92 |
| Autism spectrum disorder | 98 patients and 234 controls | Caucasian | Germany | | CC | 93 |
| Earlier age of onset HD | 419 HD patients | Caucasian | Germany | | TT | 101 |
| Variation in pre-attentive sensory memory sub-processes | 199 healthy participants | Caucasian | Germany | | CC | 106 |
| AESD | 85 patients and 184 controls | Asian | Japan | C | | 14 |
| Sustained vigilant attention after prolonged wakefulness | 45 healthy participants | Caucasian | Switzerland | C | | 110 |
| Lower effect of caffeine on attention after prolonged wakefulness | 45 healthy participants | Caucasian | Switzerland | C | | 110 |
| **rs35320474** | | | | | | |
| Self-reported anxiety after caffeine administration | 94 healthy participants | Caucasian* | USA | | Tins/Tins | 81 |
| Self-reported anxiety after d-amphetamine administration | 99 healthy participants | n.a. | USA | | Tins/Tins | 83 |
| ADHD traits | 1747 twins | Caucasian | Sweden | Tins | | 91 |
| Increased autism-specific impairment | 98 patients | Caucasian | Germany | del | | 93 |
| Migraine with aura *vs* without aura | 122 patients aura+ and 143 aura- | Caucasian | Germany | Tins | | 99 |
| Earlier age of onset HD | 419 HD patients | Caucasian | Germany | | del/del | 101 |
| AESD | 85 patients and 184 controls | Asian | Japan | del | | 14 |
| Sustained vigilant attention after prolonged wakefulness | 45 healthy participants | Caucasian | Switzerland | del | | 110 |
| Lower effect of caffeine on attention after prolonged wakefulness | 45 healthy participants | Caucasian | Switzerland | del | | 110 |

ADHD: attention deficit/hyperactivity disorder patients; AESD: Acute encephalopathy with biphasic seizures and late reduced diffusion; BS: bladder syndrome; HD: Huntington's disease; Ref: references; del: deletion; ins: insertion; n.a.: not available; *: mostly Caucasian.

## Figure legend

**Figure 1. Adenosine receptor genes in the human.**

Schematic representation of the (**A**) *ADORA1,* (**B**) *ADORA2A,* (**C**) *ADORA2B*, and (**D**) *ADORA3* genes. Exons and introns are shown as boxes and horizontal lines, respectively. Untranslated regions are indicated as grey boxes, and coding exons are marked in black. The different TSSs are indicated by black perpendicular arrows. Transcripts with dotted lines correspond to mRNAs not fully sequenced or with low levels of evidence. Dotted boxes represent the CpG islands. Hatched boxes mark the relative location of the promoter region, according to the histone code and chromatin state segmentation data from the ENCODE project.

**Figure 1. Adenosine receptor genes in the human.**

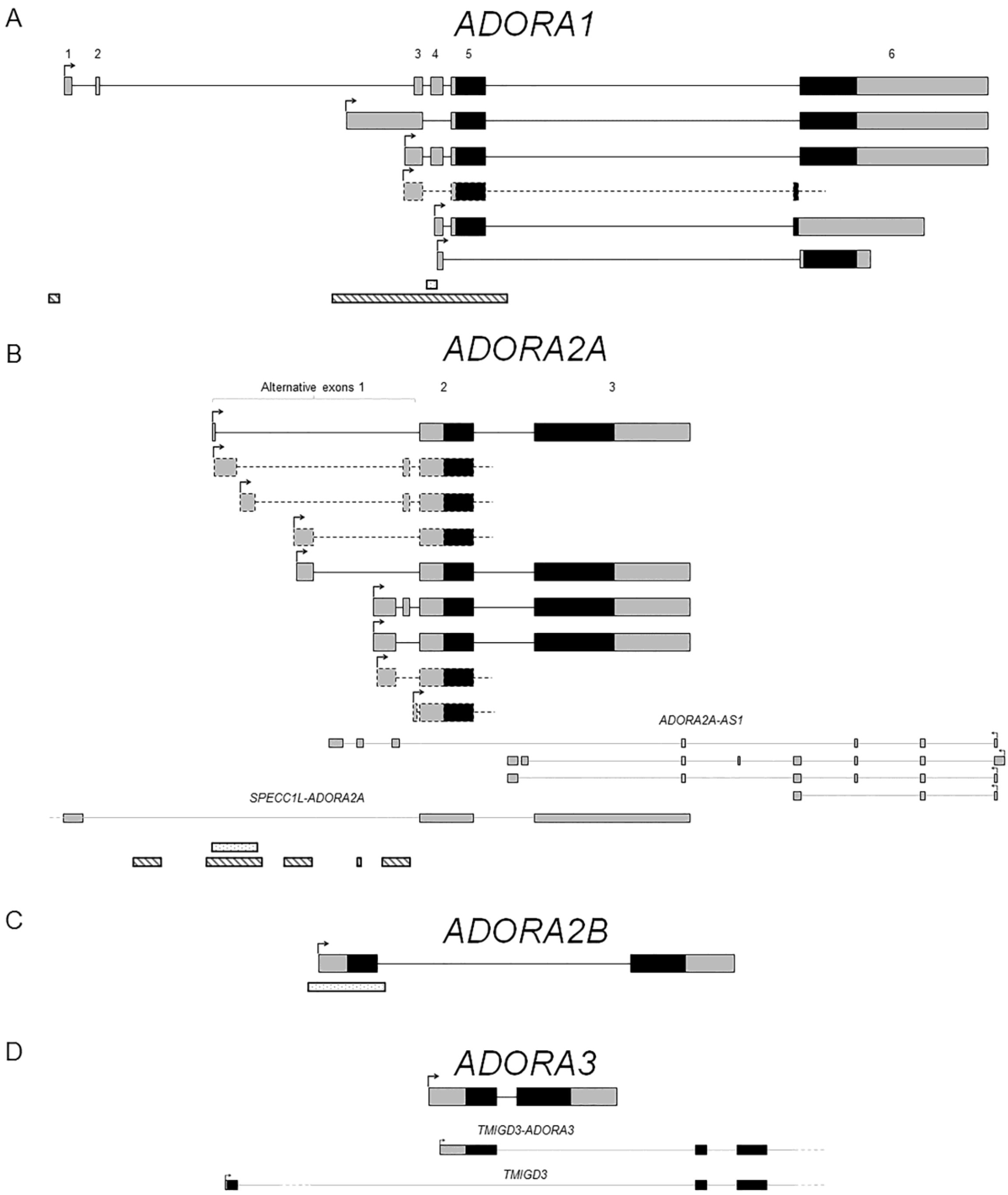

## Supplementary material

## Methods

We searched the literature published in English from MEDLINE PubMed, Science Direct, Web of Science, Google Scholar, OMIM and the GWAS catalog (https://www.ebi.ac.uk/gwas/home) until 1 April 2019. We used the search terms ("ADORA" or "Adenosine Receptor" or "ADORA1" or "RDC7" or "ADORA2A" or "A2aR" or "RDC8" or "ADORA2B" or "ADORA2" or "ADORA3" or "A3AR") AND ("Variant*" or "SNP" or "Polymorphism*" or "Genotype" or "Haplotype" or "Allel*", or "Neurology" or "Psychiatry" or "Pharmacogenetics" or "Promoter" or "Exon" or "Intron" or "Splicing" or "Translation" or "Transcript*" or "Epigen*"). Bibliographies of the publications that met these criteria and of each relevant review were also manually screened to ensure no potential articles were missed. Publications included in this review involved at least one of the following criteria: (a) Adenosine receptor gene variation implied in a human genetic trait or pathology involving the central nervous system; (b) Adenosine receptor gene cloning, structure and transcription regulation. Exclusion criteria were defined as follows: papers addressing non-neurological diseases or involving the peripheral nervous system, papers addressing only non-human genes. We excluded animal studies, conference proceedings, and editorials.